\documentclass[aps,prl,twocolumn,superscriptaddress,showpacs]{revtex4}

\usepackage{mathrsfs}
\usepackage{bm}
\usepackage{verbatim}

\usepackage{graphicx}
\usepackage{latexsym}
\usepackage{float}
\usepackage{stmaryrd}
\usepackage{amsmath}
\usepackage{amsfonts}
\usepackage{hyperref}
\usepackage{chngcntr}

\begin{document}
\title{Gapped Topological Kink States and Topological Corner States in Graphene}
\author{Yuting Yang}
\affiliation{School of Physical Science and Technology, Soochow University, Suzhou 215006, China}
\author{Ziyuan Jia}
\affiliation{School of Physical Science and Technology, Soochow University, Suzhou 215006, China}
\author{Yijia Wu}
\affiliation{International Center for Quantum Materials, Peking University, Beijing 100871, China}
\author{Zhi Hong Hang} \email{zhhang@suda.edu.cn}
\affiliation{School of Physical Science and Technology, Soochow University, Suzhou 215006, China}
\affiliation{Institute for Advanced Study, Soochow University, Suzhou 215006, China}
\author{Hua Jiang} \email{jianghuaphy@suda.edu.cn}
\affiliation{School of Physical Science and Technology, Soochow University, Suzhou 215006, China}
\affiliation{Institute for Advanced Study, Soochow University, Suzhou 215006, China}
\author{X. C. Xie}
\affiliation{International Center for Quantum Materials, Peking University, Beijing 100871, China}
\affiliation{Beijing Academy of Quantum Information Sciences, Beijing 100193, China}
\affiliation{CAS Center for Excellence in Topological Quantum Computation, University of Chinese Academy of Sciences, Beijing 100190, China}

\date{\today}
\begin{abstract}
Based on the tight-binding model calculations and photonic experimental visualization on graphene, we report the domain-wall-induced gapped topological kink states and topological corner states. In graphene, domain walls with gapless topological kink states could be induced either by sublattice symmetry breaking or by lattice deformation. We find that the coexistence of these two mechanisms will induce domain walls with gapped topological kink states. Significantly, the intersection of these two types of domain wall gives rise to topological corner state localized at the crossing point. Through the manipulation of domain walls, we show graphene not only a versatile platform supporting multiple topological corner modes in a controlled manner, but also possessing promising applications such as fabricating topological quantum dots composed of gapped topological kink states and topological corner states.
\end{abstract}

\date{\today}
\pacs{72.80.Vp, 72.10.-d, 73.20.At}
\maketitle
\textit{Introduction.---}
The gapless topological kink states have been widely investigated in domain walls (DWs) of graphene-type electronic and classical wave systems \cite{Martin,Semenoff,FZhang, Vaezi,XiaoHu1,Qiao1, Qiao2,SGCheng,JZhu2}. Generally, there are two different mechanisms of the formation of DWs. The first one is through sublattice symmetry breaking. For example, a DW separating two graphene sheets with oppositely staggered AB sublattice potential (AB-BA DW) hosts gapless topological kink states \cite{Martin,Semenoff,FZhang,Vaezi}, which has been experimentally demonstrated in graphene \cite{FWang,JZhu1,LinHe1}, sonic and photonic graphene {\textit{etc}} \cite{ZLiu,Dong,  liping,ZHHang2, Zhang,Noh}. The second mechanism is lattice deformation, where gapless topological kink states localized along the interface between two graphene sheets with reversed  $\sqrt{3} \times \sqrt{3}$ lattice deformation ($\sqrt{3} \times \sqrt{3}$ DW) \cite{XiaoHu1}. Such a mechanism has also been experimentally visualized in various clasical wave graphene systems \cite{YFCheng,YCheng,ZHHang,HChen}. Nevertheless, the topological states in the coexistence of these two mechanisms still remain to be investigated.

Higher-order topological insulators (HOTIs) \cite{Hughes,Oppen,ZDSong, Neupert,Ezawa,Garcia, Peterson,Thomale,Rechtsman, Khanikaev,Zhang2,JHJiang,HSChen} have been attracting intensive attentions for its unique bulk-boundary correspondence. The non-trivial gapped edge states and corner states in two-dimensional (2D) HOTIs are protected by the bulk gap, which have been experimentally observed in various classical systems with square or kagome lattices \cite{Hughes, Garcia,Peterson, Thomale,Rechtsman,Khanikaev, Zhang2,JHJiang, HSChen}. It is natural to ask that whether these topological states can also be realized in graphene systems. Considering the manipulability of both AB-BA and $\sqrt{3} \times \sqrt{3}$ DWs and the rich topological phase diagram, graphene systems may open a new way integrating multiple topological states on a single chip.

\begin{figure}
\includegraphics[width=8cm]{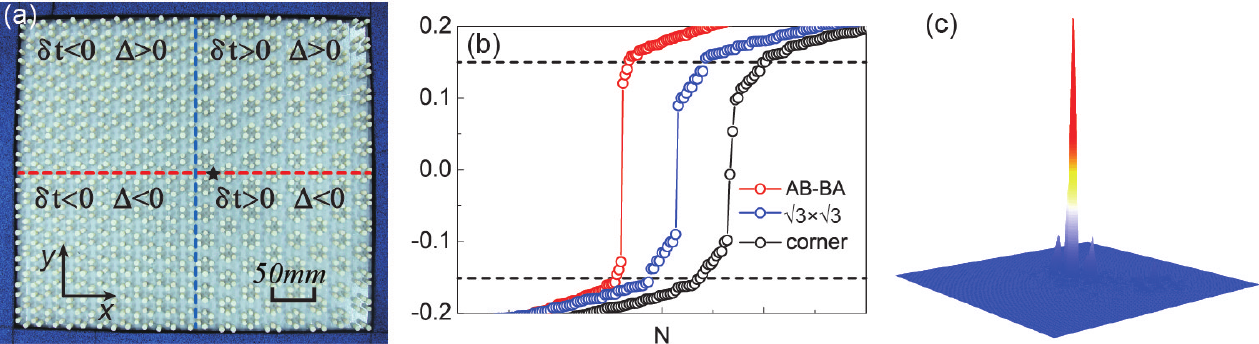}
\caption{(Color online)
(a) Photograph of the fabricated photonic graphene device. Red (blue) line denotes the AB-BA ($\sqrt{3} \times \sqrt{3}$) DW. (b) Numerical results of the eigen-energy spectrum of the  device with AB-BA DW only, $\sqrt{3} \times \sqrt{3}$ DW only, and two types of DWs intersected, respectively. The two dashed lines indicate the bulk gap. (c) Measured electric field distribution of the topological corner state at 7.3590~GHz.}
\label{Fig1}
\end{figure}

In this Letter, through theoretical analysis and photonic crystal (PC) experiments, we study the DW-induced topological states in graphene in the coexistence of sublattice symmetry breaking and lattice deformation. Introducing lattice deformation to the AB-BA DW [red lines in Fig. \ref{Fig1}(a), (b)] or breaking sublattice symmetry of a $\sqrt{3} \times \sqrt{3}$ DW [blue lines in Fig. \ref{Fig1}(a), (b)] will both induce gapped topological kink states. Significantly, the crossing of AB-BA and $\sqrt{3} \times \sqrt{3}$ DWs [Fig. \ref{Fig1}(a)] gives rise to topological corner states [black dot in Fig. \ref{Fig1}(b)] localized at the intersection [Fig. \ref{Fig1}(c)]. Moreover, we show the manipulability of the DWs, thus demonstrating graphene a versatile platform supporting multiple topological corner states with tunable interactions. Finally, we propose a scheme fabricating topological quantum dots (TQDs) possessing topological corner states, gapless and gapped topological kink states,  implying promising applications in mesoscopic physics.

\begin{figure*}
\includegraphics [width=13.5 cm, viewport=0 0 450 186, clip]{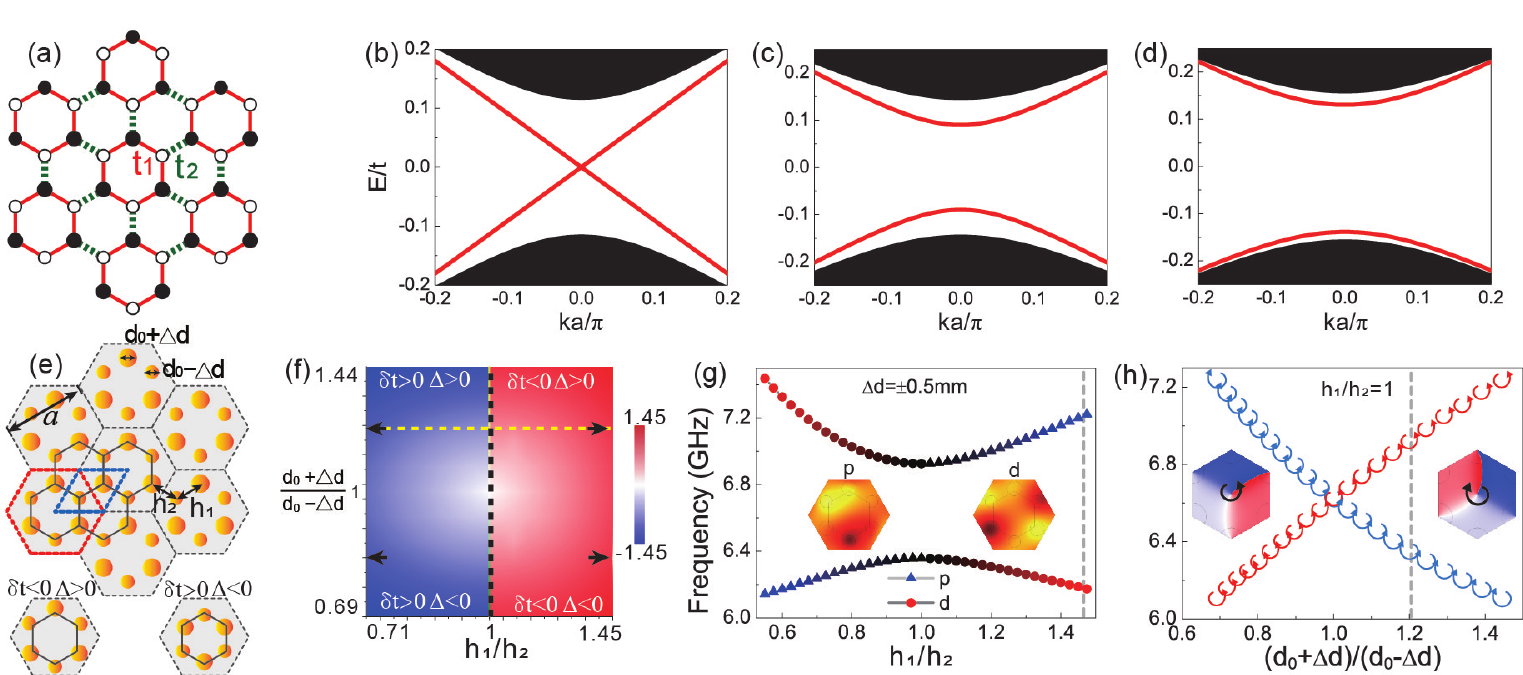}
\caption{(Color online) (a) Illustration of the tight-binding model. The red solid (green dashed) bond denotes modulated hopping strength $t_1=t+\delta t$ ($t_2=t-\delta t$). The solid (hollow) circles represent lattice site with onsite potential $\Delta$ ($-\Delta$). DW is the line across which $\Delta$ or $\delta t$ changing its sign. (b)-(d) 1D band spectra of DW structure with zigzag interface. The parameters are $(\delta t, \Delta)$=(0.06t, 0)/(-0.06t, 0) in (b); (0.06t, 0.09t)/(-0.06t, 0.09t) in (c); and (0.0675t, 0.09t)/(0.0675t, -0.09t) in (d), respectively. (e) Schematics of the experimental setup utilizing PCs. Nonzero $\Delta$ and $\delta t$ are realized by manipulating the cylinder diameter $d_0 \pm \Delta d$ on A/B sublattice and the separation between neighboring cylinders $h_1, h_2$, respectively. (f) Phase diagram in the parameter space spanned by $\frac{d_0+\Delta d}{d_0-\Delta d}$ and $\frac{h_1}{h_2}$. (g), (h) 1D phase transition occurs along the yellow (black) dashed line in (f) by (g) lattice deformation [(h) sublattice symmetry breaking].}
\label{Fig2}
\end{figure*}

\textit{Theoretical model.---}
As shown in Fig. \ref{Fig2} (a), we begin with a modified graphene tight-binding model $H= -\sum_{\langle i j \rangle} t_{ij} c_{i}^{\dag}c_{j}+ \sum_{i} U_{i} c_{i}^{\dag} c_{i}$, where $c_{i}^{\dag}$ is the creation operator on site $i$. The $\sqrt{3} \times \sqrt{3}$ lattice deformation is considered by modulating the intra (inter)-unit-cell hopping strength $t_{ij}$ as $t_1 =t +\delta t$ ($t_2=t-\delta t$). Noticing that the primitive unit cell is now enlarged and contains six sites. The sublattice symmetry breaking is included by adding staggered potential $U_{i}=\Delta$ ($-\Delta$) on the A (B) sublattice. When $t_1$ and $t_2$ are exchanged (reversed $\delta t $) across a line, a $\sqrt{3} \times \sqrt{3}$ DW is formed. Similarly, a AB-BA DW is formed when $\Delta$ changes its sign \cite{Semenoff}.

We first study the relationship bewtween topological states in these two types of DWs. Figure \ref{Fig2}(b) plots the 1D energy spectrum for the $\sqrt{3} \times \sqrt{3}$  DW. Counter-propagating gapless states inside the bulk gap \cite{XiaoHu2} are similar to the topological kink states in the AB-BA DW \cite{Semenoff,SGCheng,Yao}. To comprehensively understand the similarity, we derive the low-energy effective Hamiltonian of $H$ following Ref. \cite{XiaoHu2, Supp,XiaoHu3}

\begin{eqnarray}
\mathcal{H}_{\rm eff} (k) = \left(
                          \begin{array}{cccc}
                             t_1 -t_2 & -\hbar \upsilon k_{-} & 0 & i \Delta \\
                             -\hbar \upsilon k_{+} & t_2-t_1 & i \Delta & 0 \\
                            0 & -i \Delta & t_1-t_2 &   \hbar \upsilon k_{+} \\
                             -i \Delta & 0 &   \hbar \upsilon k_{-}  & t_2-t_1 \\
                          \end{array}
                        \right)
, \label{EQ1}
\end{eqnarray}
where $k_{\pm}=k_y \pm i k_x$ and $\upsilon$ denotes the Fermi velocity. Equation (\ref{EQ1}) with $\Delta=0$ is analogous to the Hamiltonian of sublattice-symmetry-breaking graphene ($\Delta \neq 0, \delta t =0$) in the basis of $\{ K, K' \} \otimes \{A , B \}$. They are topologically equivalent for $t_1-t_2$ playing the role of $\Delta$. Therefore, the gapless states in the $\sqrt{3} \times \sqrt{3}$ and AB-BA DWs have the same topology origin, where $t_1-t_2$ and $\Delta$ play the role of the mass term, respectively. The sign change of the mass term across the DW leads to the bulk band inversion, which protects the gapless topological kink states.

However, $t_1-t_2$ and $\Delta$ possesses different positions in Eq. (\ref{EQ1}), and the bulk gap is $E_g = 2 \sqrt{(t_1-t_2)^2+ \Delta^2 }$. These observations manifest that the mass terms $t_1-t_2$ and $\Delta$ are orthogonal to each other. It indicates that even though one mass term changes its sign across the domain wall, the topological kink state are still gapped due to the presence of another mass term. As plotted in Fig. \ref{Fig2}(c) [\ref{Fig2}(d)], there is a $2|\Delta|$ ($2|t_1-t_2|$) gap in the topological kink state of $\sqrt{3} \times \sqrt{3}$ (AB-BA) DW by introducing lattice deformation (breaking sublattice symmetry).

We further study the intersection of these two types of DWs. For convenience, we assume the AB-BA ($\sqrt{3} \times \sqrt{3}$) DW is along the $y$ ($x$) direction [Fig. \ref{Fig1}(a)]. In both the left and right side of the AB-BA DW, the gapped topological kink state can be described as $\hbar \upsilon k_x \sigma_z +m_t \sigma_x$, where $m_t=t_1-t_2$ is the mass term by lattice deformation. However, $m_t$ changes its sign due to the $\sqrt{3} \times \sqrt{3}$ DW, leading to a 1D band inversion. Consequently, a topological corner state emerges around the crossing point. Fig. \ref{Fig1}(b) exhibits their eigen-energy spectrum, a nearly-zero-energy state inside the gap of the topological kink states is observed only in the presence of crossing of AB-BA and $\sqrt{3} \times \sqrt{3}$ DW, indicating the existence of the topological corner state.

\textit{Experimental observation.---}
A photonic graphene is experimentally designed and fabricated by assembling alumina ${\rm Al_2O_3}$ cylinders with relative permittivity $\epsilon =7.5$, as illustrated in Fig. \ref{Fig2}(e). The hexagon unit cell (honeycomb sublattice) is indicated in red (blue). The hopping strength $t_1$ ($t_2$) in the tight-binding model is controlled by the intra (inter)-unit-cell separation between neighbouring cylinders $h_1$ ($h_2$) \cite{ZHHang}. The sublattice potential $\Delta$ is modulated by the diameters of the cylinder. The predictions of the tight-binding model can be easily checked in PCs for both $h_{1/2}$ and $d_0$ in the centimeter-scale [Fig. \ref{Fig1}(a)].

We firstly calculate the phase diagram with different parameters $h_1$, $h_2$, $d_0$ and $\Delta d$ [Fig. \ref{Fig2}(f)]. Here $d_0$ is set as $5.5$~mm and the lattice constant $a=2h_1+h_2=17\sqrt{3}$~mm. There are two doubly degenerate modes at the center of the Brillouin zone for each PC \cite{Supp}. The dipole (quadruple)-like symmetry can be a photonic analogue of the $p$ ($d$)-electron orbit. As shown in Fig. \ref{Fig2}(g), with the increase of the ratio $h_1/h_2$, the frequency of the dipole-like modes (blue triangles) exceeds the quadruple-like one (red circles), indicating the band inversion induced by the lattice deformation. Their frequency difference (in the unit of GHz) is shown by the color in Fig. \ref{Fig2}(f). When the cylinder diameter changes [Fig. \ref{Fig2}(h)], the eigen-frequency of the counter-clockwise magnetic field exceeds the clockwise counterpart, indicating 1D phase transition by staggered A/B sublattice \cite{ZHHang2}. Therefore the phase diagram can be separated into four quadrants in the coexistence of two different 1D phase transitions.

\begin{figure*}
\includegraphics[width=13.5 cm]{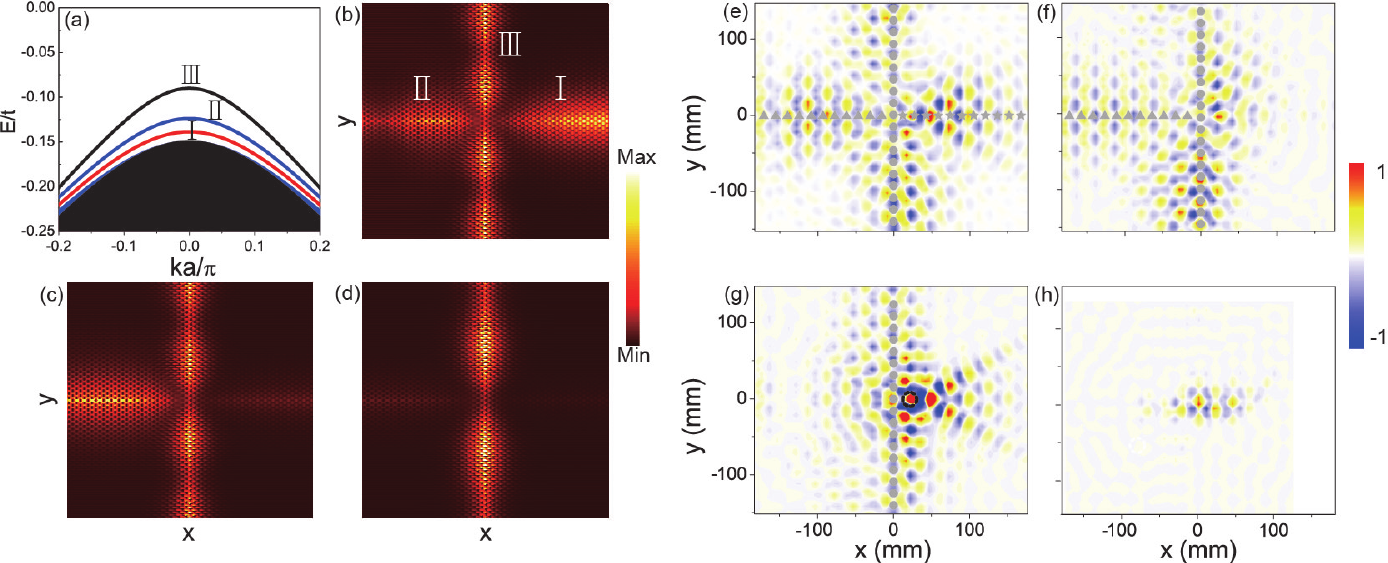}
\caption{(Color online) (a) Energy spectrum for the device with DWs I, II, and III. The parameters of the four quadrants are $(\delta t, \Delta)$ =(0.0675t, 0.09t), (-0.06t, 0.09t), (-0.06t, -0.09t), and (0.0675t, -0.09t), respectively. (b)-(d) LDOS distributions at typical energies $E = 0.14t, 0.12t, 0.10t$, respectively. (e)-(h) Experimentally measured electric field distributions at frequencies $f=7.1545, 7.1895, 7.2005,$ and $7.4115$~GHz, respectively. (h) Topological corner mode survives in the presence of disorders.}
\label{Fig3}
\end{figure*}

We choose one set of parameters from each quadrant [denoted by arrows in Fig. \ref{Fig2}(f)] to construct the intersection of the two types of DWs. The photograph of the experimental setup is shown in Fig. \ref{Fig1}(a), where the blue line denotes the $\sqrt{3} \times \sqrt{3}$ DW with $h_1= 19/\sqrt{3}$~mm ($h_1=13.5/\sqrt{3}$~mm) on its left (right) side. The cylinders at site A (B) are in the diameter of $5$~mm ($6$~mm) and the locations of A/B sites are swapped crossing the AB-BA DW [red line in Fig. \ref{Fig1}(a)]. These cylinders' height is $8$mm and they are assembled inside a parallel-plate waveguide (the upper metallic plate is not shown), thus the polarization of the electric field is parallel to the cylinders. In principle, there could be four different DWs while in order to facilitate the sample preparation, the parameters we choose give rise to three regions [Fig. \ref{Fig3}(b)] and their corresponding energy spectra using tight-binding model are shown in Fig. \ref{Fig3}(a). Though the gapped topological kink states exist in all three regions {\rm I}, {\rm II} and {\rm III}, their energy gaps are different. The local density of states (LDOS) distributions at typical Fermi energies obtained by lattice Green's function are plotted in Fig. \ref{Fig3}(b)-(d) \cite{Datta}, where three patterns of gapped topological kink states are shown with the increase of the Fermi energy.

Through exciting the photonic graphene by a monopole antenna [denoted by star in Fig. \ref{Fig1}(a)], time-harmonic electric field distributions at different frequencies can be measured [Fig. \ref{Fig3}(e)-(h)]. In case of the frequency $f=7.1545$~GHz slightly higher than the top of the bulk photonic bands [Fig. \ref{Fig3}(b)], measured electric field is localized around all of the DWs, indicating topological kink states in all regions I, II, and III [Fig. \ref{Fig3}(e)]. By increasing the frequency to $f=7.1895$~GHz, only the states near region II and III [Fig. \ref{Fig3}(f)] are excited. Such observations are in consistent with the LDOS distribution in Fig. \ref{Fig3}(c) where the Fermi energy only crosses the topological kink bands II and III, showing a strong evidence of the existence of gapped topological kink states in region I. In case that the frequency is increased to $f=7.2005$~GHz, the gapped topological kink states in region I and II [Fig. \ref{Fig3}(d), (g)] are manifested. Significantly, when the frequency increases to $f=7.3590$~GHz, only the states around $x=y=0$ are excited [Fig. \ref{Fig1}(c)]. The spatial exponential decay of the corresponding electric field intensity (see Supplemental Material for more detailed analysis \cite{Supp}) clearly characterizes the topological corner modes. Moreover, the topological corner state is robust against disorder effect introduced by random dislocations of ${\rm Al_2O_3}$ cylinders \cite{Supp} [Fig. \ref{Fig3}(h)].

\begin{figure}
\includegraphics [width=8cm]{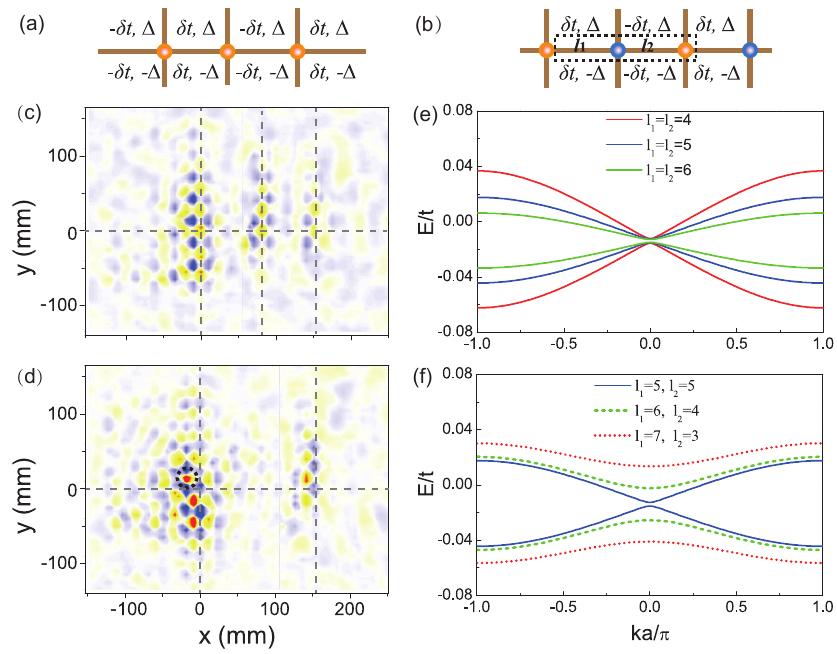}
\caption{(Color online) (a) Schematic diagram of the allignment of DWs with three topological corner modes. (b) Illustration of SSH chain. For each unit, there are two topological corner modes with bond length $l_1, l_2$. (c),(d) Experimentally measured field distributions with topological corner modes coupling at $f=7.246$, and $7.260$~GHz, respectively. (e),(f) Energy spectrum for (b) in the presence of different $l_1,l_2$'s ($\Delta =0.18t$, $\delta t =0.06t$).}
\label{Fig4}
\end{figure}

\textit{Applications.---}
In addition to the simple HOTI possessing single topological corner mode, one can create a number of topological corner modes and manipulate their interactions by arranging the spatial positions of these DWs. A device with three topological corner modes is illustrated in Fig. \ref{Fig4}(a), which can be described by Hamiltonian $\mathcal{H}_1= -\tau a_1^{\dag}a_2 -\tau a_2^{\dag}a_3 + h.c.$. Such device's spatial distributions of the electric field at frequencies $f=7.246$ and $7.260$~GHz are measured. In Fig. \ref{Fig4}(c), three peaks of the electric field are observed, corresponding to the eigen-vector $[\frac{1}{2},-\frac{1}{\sqrt{2}},\frac{1}{2}]^{T}$ for the lowest eigen-energy of $\mathcal{H}_1$. Contrastly, the eigen-vector at medium eigen-energy is $[\frac{1}{\sqrt{2}}, 0, \frac{1}{\sqrt{2}}]^{T}$, therefore the electric field vanishes in the central region as confirmed in Fig. \ref{Fig4}(d). Due to the leakage from excitation antenna [dashed circle in Fig. \ref{Fig4}(d)], the field near the leftmost corner mode is excessively strong. One can also periodically arrange the DWs to construct a Su-Schrieffer-Heeger (SSH) chain \cite{SSH} by topological corner modes [Fig. \ref{Fig4}(b)]. There are two topological corner modes with bond length $l_1$ and $l_2$ in a unit cell. Fig. \ref{Fig4}(e), (f) show simulated energy spectrum. Increasing $l_1$ and $l_2$ simultaneously will decrease both the inter- and intra-unit-cell coupling strength, therefore the band width also decreases [Fig. \ref{Fig4}(e)]. In case that $l_1>l_2$, the inter-unit-cell coupling strength is stronger than the intra-unit-cell one. A band gap emerges as shown in Fig. \ref{Fig4}(f) and increases with the difference between $l_1$ and $l_2$. Since topological corner modes can be used to construct different tight-binding models with tunable parameters, such platform bears similarities to the optical lattice in cold atom \cite{Morsch} and is much more convenient in sample fabrication and measurement.

\begin{figure}
\includegraphics[width=8cm]{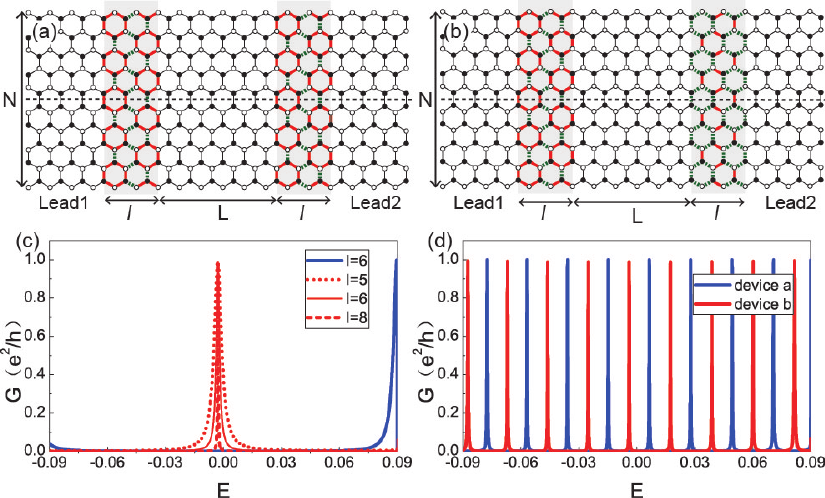}
\caption{(Color online) (a),(b) Illustration of TQD devices, which includes QD (region with length $L$), barriers (region with length $l$) and semi-infinite leads (here $l=1$, $L=2$, and $N=4$). The $\sqrt{3} \times \sqrt{3}$ domain wall is absent (present) in (a) [(b)]. (c),(d) Conductance $G$ \textit{vs.} Fermi energy $E$. The blue (red) lines represent the device in (a) [(b)]. In both two leads and QD, $(\delta t, \Delta) =(0, 0.09t)$. In two barriers, $(\delta t, \Delta )=(0.06t, 0.09t)$. The other parameters are (c) $L=2$, and (d) $L=40$.}
\label{Fig5}
\end{figure}

The coexistence of multiple kinds of topological states in such graphene system, especially in photonic graphene, provides the possibility of fabricating unique topological devices. We propose two TQDs as illustrated in Fig. \ref{Fig5}(a), (b) by combining different topological states. For both two devices, a AB-BA DW is formed across the dashed line, while $\sqrt{3} \times \sqrt{3}$ lattice deformation is present only in the filled region [Fig. \ref{Fig5}(a), (b)]. The gapped topological kink states exist in such regions, acting as two barriers separating the central region from the two leads with gapless topological kink states. Due to the finite-size confinement, the bound states are formed in the central region \cite{JJZhou}. Fig. \ref{Fig5}(c), (d) show the conductance $G$ as function of Fermi energy $E$ of these two TQDs. The key feature is that $G$ shows resonant tunneling behavior when $E$ approaches the energy of the bound state  $\varepsilon$ \cite{Jauho}. For a very small $L$ [Fig. \ref{Fig5}(c)], the devices that including two leads is similar to the system in  Fig. \ref{Fig1}(a). A sharp peak of $G$ emerges around $E=0$ only for the device in Fig. \ref{Fig5}(b). These results agree with Fig. \ref{Fig1}(b), where zero-energy topological corner state is localized at the crossing of AB-BA and $\sqrt{3} \times \sqrt{3}$ DWs. In the case of large $L$, a half-period shift of $G$ peaks between two devices is observed, because the $\sqrt{3} \times \sqrt{3}$ DW induces a special boundary condition and changes the bound energy equation from $e^{i\frac{2 \varepsilon }{\hbar \upsilon}L}=\frac{ \varepsilon+ i \sqrt{|m_t|^2- \varepsilon ^2}}{\varepsilon - i \sqrt{|m_t|^2- \varepsilon ^2}}$ [Fig. \ref{Fig5}(a)] to $e^{i(\frac{2 \varepsilon }{\hbar \upsilon}L + \pi)}=\frac{ \varepsilon + i \sqrt{|m_t|^2- \varepsilon ^2}}{\varepsilon - i \sqrt{|m_t|^2- \varepsilon ^2}}$ [Fig. \ref{Fig5}(b)]. Since $\varepsilon=0$ is always the solution of the latter equation, the bound energy equation also verifies that such topological corner states are examples of the celebrated Jackiw-Rebbi zero-mode \cite{Jackiw}. The detailed derivations of $\varepsilon$ for the two TQDs and the comparison to normal QDs are summarized in the Supplemental Materials\cite{Supp}. Considering the key role that QD plays in mesoscopic device, this proposal may opens up a new path simulating mesoscopic physics in PCs.

\textit{Summary.---} 
DW-induced gapped topological kink states and topological corner states are theoretically predicted and experimentally observed in graphene. The gapped topological kink states in DWs are protected by the orthogonality between the two mass terms induced by sublattice symmetry breaking and lattice deformation. Besides, the topological corner states emerge in the presence of intersection between these two types of DWs. Furthermore, through the manipulation of DWs, we also illuminate two applications by integrating multiple topological states in a single device. A versatile platform is proposed where the number of topological corner modes and their interactions are tunable. The advantages of applying mesoscopic physics concepts in photonic graphene are also demonstrated by constructing TQDs where gapped topological kink states and topological corner states coexist.

\textit{Acknowledgements.---} We are grateful to H. M. Weng and H. W. Liu for helpful discussion. This work was supported by NSFC under Grants Nos. 11534001, 11574226, 11822407 and 11874274, NSF of Jiangsu Province under Grants Nos. BK2016007 and BK20170058 and a Project Funded by the Priority Academic Program Development of Jiangsu Higher Education Institutions (PAPD).

\pagebreak

\counterwithin{equation}{section} 

\counterwithin{figure}{section} 

\setcounter{equation}{0}
\setcounter{figure}{0}
\setcounter{table}{0}
\setcounter{page}{1}
\makeatletter
\renewcommand{\theequation}{S\arabic{equation}}
\renewcommand{\thefigure}{S\arabic{figure}}
\renewcommand{\bibnumfmt}[1]{[S#1]}
\renewcommand{\citenumfont}[1]{S#1}

\begin{widetext}

\begin{center}
\textbf{Supplementary Material for ``Gapped Topological Kink States and Topological Corner States in Graphene"}
\end{center}

\begin{center}
Yuting Yang,$^1$ Ziyuan Jia,$^1$ Yijia Wu,$^2$ Zhi Hong Hang,$^{1,3,*}$ Hua Jiang,$^{1,3,\dag}$ and X. C. Xie$^{2,4,5}$
\end{center}
\begin{center}
$^1$~School of Physical Science and Technology,  Soochow University, Suzhou 215006, China

$^2$~International Center for Quantum Materials, Peking University, Beijing 100871, China

$^3$~Institute for Advanced Study, Soochow University, Suzhou 215006, China

$^4$~Beijing Academy of Quantum Information Sciences, Beijing 100193, China

$^5$~CAS Center for Excellence in Topological Quantum Computation, University of Chinese Academy of Sciences, Beijing 100190, China
\end{center}


\section{Low-energy effective Hamiltonian description of topological states}
In this section, we derive the low-energy effective Hamiltonian to analyze the topological states in the studied systems \cite{XiaoHu2}. There are six orbits in the representation basis since each unit cell has six lattice sites. By choosing the primitive vectors $a_1= a (0, 1)$ and $a_2 =a (\frac{\sqrt{3}}{2},-\frac{1}{2})$ with $a=2h_1+h_2$, the momentum space Hamiltonian in  the frame of such six orbits basis reads

\begin{eqnarray}
\mathcal{H} (k)=\left(  \begin{array}{cccccc}
                                      \Delta & -t_1 & 0 & -t_{2} e^{ik \cdot a_1 } & 0 & -t_1 \\
                                      -t_1 & -\Delta & -t_1 & 0 & -t_{2} e^{ik\cdot(a_1 +a_2)}  & 0 \\
                                      0 & -t_1 & \Delta & -t_1 & 0 & -t_{2} e^{ik \cdot a_2} \\
                                      -t_{2} e^{-ik \cdot a_1} & 0 & -t_1 & -\Delta & -t_1 & 0 \\
                                      0 & -t_{2} e^{-ik \cdot(a_1+a_2)} & 0 & -t_1 & \Delta & -t_1 \\
                                      -t_1 & 0 & -t_{2} e^{-ik \cdot a_2} & 0 & -t_1 & -\Delta \\
                                    \end{array}
                                  \right). \label{EQA1}
\end{eqnarray}

\noindent Around the Brillouin zone center $\Gamma$ point, Eq. (\ref{EQA1}) reduces:

\begin{eqnarray}
\mathcal{H} (k) \approx - \left(  \begin{array}{cccccc}
                                      -\Delta & t_1 & 0 & t_{2} (1+ik \cdot a_1 ) & 0 & t_1 \\
                                      t_1 & \Delta & t_1 & 0 & t_{2} [1+ik\cdot(a_1 +a_2)]  & 0 \\
                                      0 & t_1 & -\Delta & t_1 & 0 & t_{2} (1+ik \cdot a_2) \\
                                      t_{2} (1-ik \cdot a_1) & 0 & t_1 & \Delta & t_1 & 0 \\
                                      0 & t_{2} [1-ik \cdot(a_1+a_2)] & 0 & t_1 & -\Delta & t_1 \\
                                      t_1 & 0 & t_{2} (1-ik \cdot a_2) & 0 & t_1 & \Delta \\
                                    \end{array}
                                  \right). \label{EQA2}
\end{eqnarray}

\noindent Considering the condition that both $t_{1,2}$ approach $t$, and $\Delta \ll t$, we perform a unitary transformation

\begin{eqnarray}
U =  \frac{1}{\sqrt{6}}{\rm \left(  \begin{array}{cccccc}
                                      e^{i\frac{\pi}{2}} & e^{i \pi} &  e^{i\frac{3\pi}{2}} & e^{i \pi} & 1 & 1\\
                                       e^{i\frac{7\pi}{6}} & e^{i \frac{4\pi}{3}} &  e^{i\frac{5\pi}{6}} & e^{i \frac{2\pi}{3}} & -1 & 1 \\
                                      e^{i\frac{11\pi}{6}} & e^{i \frac{5\pi}{3}} &  e^{i\frac{\pi}{6}} & e^{i \frac{\pi}{3}} & 1 & 1  \\
                                       e^{i\frac{\pi}{2}} & e^{i 2\pi} &  e^{i\frac{3\pi}{2}} & e^{i 2\pi} & -1 & 1 \\
                                       e^{i\frac{7\pi}{6}} & e^{i \frac{ \pi}{3}} &  e^{i\frac{5\pi}{6}} & e^{i \frac{5\pi}{3}} & 1 & 1 \\
                                       e^{i\frac{11\pi}{6}} & e^{i \frac{2\pi}{3}} &  e^{i\frac{\pi}{6}} & e^{i \frac{4\pi}{3}} & -1 & 1\\
                                    \end{array}
                                  \right) }\label{EQA3}
\end{eqnarray}

\noindent and obtain the Hamiltonian in a new basis as

\begin{eqnarray}
\mathcal{H}_1 (k) &=& U^{\dag} \mathcal{H} (k) U  \nonumber\\
&=&  \left(  \begin{array}{cccccc}
                                      t_1 -t_2 & -\hbar \upsilon (k_y - i k_x) & 0 & i \Delta  &  \hbar \upsilon (k_y + i k_x) & 0 \\
                                      -\hbar \upsilon (k_y + i k_x) & t_2-t_1 & i \Delta & 0 &  0  &  \hbar \upsilon (k_x + i k_y) \\
                                      0 & -i \Delta & t_1-t_2 &   \hbar \upsilon (k_y + i k_x) & -\hbar \upsilon (k_y - i k_x) & 0 \\
                                      -i \Delta & 0 &   \hbar \upsilon (k_y - i k_x)  & t_2-t_1 & 0 & -\hbar \upsilon (k_x - i k_y) \\
                                      \hbar \upsilon (k_y - i k_x) & 0 & -\hbar \upsilon (k_y + i k_x) & 0 & 2t_1+t_2 & \Delta  \\
                                      0 & \hbar \upsilon (k_x - i k_y) & 0 & -\hbar \upsilon (k_x + i k_y) & \Delta & -2t_1 - t_2 \\
                                    \end{array}
                                  \right)
                                  , \label{EQA4}
\end{eqnarray}

\noindent where $\hbar \upsilon =\frac{ta}{2}$. Because $|t_1-t_2|$ and $|\Delta|$ are much smaller than $|2t_1+t_2|$, the high energy part of $\mathcal{H}_1 (k) $ by the last two orbits of the basis can be neglected.  Thus, we obtain the final effective Hamiltonian to the first-order in $k$,

\begin{eqnarray}
\mathcal{H}_{\rm eff} (k) = \left(
                          \begin{array}{cccc}
                             t_1 -t_2 & -\hbar \upsilon (k_y - i k_x) & 0 & i \Delta \\
                             -\hbar \upsilon (k_y + i k_x) & t_2-t_1 & i \Delta & 0 \\
                            0 & -i \Delta & t_1-t_2 &   \hbar \upsilon (k_y + i k_x) \\
                             -i \Delta & 0 &   \hbar \upsilon (k_y - i k_x)  & t_2-t_1 \\
                          \end{array}
                        \right)
, \label{EQA5}
\end{eqnarray}

\noindent Before analyzing $\mathcal{H}_{\rm eff} (k)$, we revisit the previous studies of topological kink states in AB-BA domain wall where sublattice symmetry is broken. In the basis of $\{ K, K' \} \otimes \{A, B \}$, such low-energy Hamiltonian reads:

\begin{eqnarray}
\mathcal{H}_{AB} (k) = \left(
                          \begin{array}{cccc}
                            \Delta & -\hbar \upsilon (k_x - i k_y) & 0 & 0 \\
                            - \hbar \upsilon (k_x + i k_y) & -\Delta & 0 & 0 \\
                            0 & 0 & \Delta &  - \hbar \upsilon (k_x + i k_y) \\
                             0 & 0 &  - \hbar \upsilon (k_x - i k_y)  & -\Delta \\
                          \end{array}
                        \right)
, \label{EQA6}
\end{eqnarray}

\noindent Equation (\ref{EQA5}) with $\Delta =0$ describes the state with $\sqrt{3} \times \sqrt{3}$ lattice deformation only. We compare it with  Eq. (\ref{EQA6}). The exchange between momentum components $k_x$ and $k_y$ as well as the reversal of the Fermi velocity $\upsilon$ will not alter the topological nature. Therefore, though the basis of these two equations are different, $\mathcal{H}_{\rm eff} (k)$ with $\Delta=0$ is topologically equivalent to $\mathcal{H}_{AB} (k)$ except the mass term changing from the hopping energy difference $t_1-t_2$ to the staggered potential $\Delta$.  Therefore, one can regard that the gapless topological kink state both in the AB-BA domain wall and in the lattice deformation domain wall have the same physical origin, where either $t_1-t_2$ or $\Delta$ plays the role of mass term, respectively.

It is also worth to note that the mass term $t_1-t_2$ and $\Delta$ are not identical. As described in the Eq. (\ref{EQA6}), $t_1-t_2$ and $\Delta$ terms are not in the same position of the Hamiltonian. Actually, one can obtain an energy gap $E_g=2 \sqrt{(t_1-t_2)^2 +\Delta^2}$, which means the mass terms $t_1-t_2$ and $\Delta$
are orthogonal to each other. This relationship leads to the exotic gapped topological kink states and topological corner states, which are carefully studied in  the main text.

\section{Numerical results of topological corner states}
\begin{figure}[h]
\centering
\includegraphics[width=14cm]{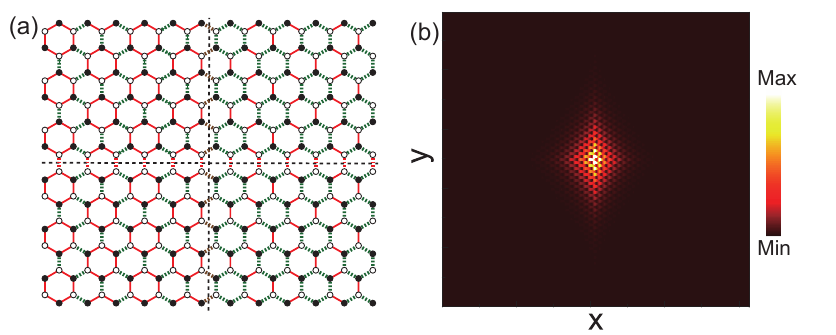}
\caption{(Color online) (a) Schematic plot of the tight-binding model description of the experimental device. The hopping energy for red (blue) bond is $t+\delta t$ ($t-\delta t$). The solid (hollow) site has on-site energy $\Delta$ ($-\Delta$). In the  four quadrants, $(\delta t, \Delta)$ are (0.0675t,~0.09t),~(-0.06t,~0.09t),~(-0.06t,~-0.09t),~(0.0675t,~-0.09t), respectively. (b) The spatial distribution of LDOS at the eigen-energy of the topological corner state, which is consistent with the experimental observation in Fig. 1(c) in the main text.}\label{FigR1}
\end{figure}

In this section, we provide the numerical results of the topological corner state. Figure S1(a) illustrates the tight-binding model description of experimental device as shown in Fig. 1(a) in the main text. Fig. S1(b) plots the spatial distribution of the LDOS of the topological corner state. Fig. S2 simulates the spatial distribution of LDOS for the device with three topological modes as illustrated in Fig. 4(a) in the main text. One can find that the numerical results in Fig. S2(a), (b) agree well with  the experimental observations in Fig. 4(c), (d) in the main text, respectively.

\begin{figure}[h]
\centering
\includegraphics[width=14cm]{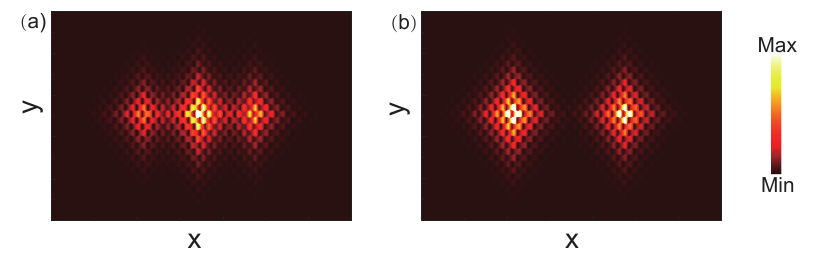}
\caption{(Color online) The spatial distribution of LDOS for the device with three topological corner modes at (a) lower eigen-energy, and (b) medium eigen-energy. The parameters are $(\delta t, \Delta)=(0.06t, 0.18t)$. (a),(b) are corresponding to the experimental results shown in Fig. 4(c), (d) of the main text, respectively.}\label{FigR2}
\end{figure}

\section{One-dimensional energy spectrum  for $\sqrt{3} \times \sqrt{3}$ domain wall  with armchair type interface}

In this section, we compare the one-dimensional (1D) energy spectrum for $\sqrt{3} \times \sqrt{3}$ domain wall with different types of interface. In the main text, the gapless topological kink state is observed for $\sqrt{3} \times \sqrt{3}$ domain wall in the presence of zigzag-type interface. When the interface is in the armchair-type, a minor gap is obtained [e.g. 0.004t compare to the bulk gap 0.24t, as plotted in Fig. S3(b)]. Such a difference may originates from the inter-valley scattering, and has also been reported in AB-BA domain-wall-induced topological kink states in the presence of armchair-type interface \cite{Qiao}. The same behavior of the topological states for AB-BA and $\sqrt{3} \times \sqrt{3}$ domain wall in the presence of different types of interface is consistent with the conclusion that the hopping energy difference $t_1-t_2$ and the staggered potential $\Delta$ play the same role in the band topology protecting the topological states.

\begin{figure}[h]
\centering
\includegraphics[width=12cm]{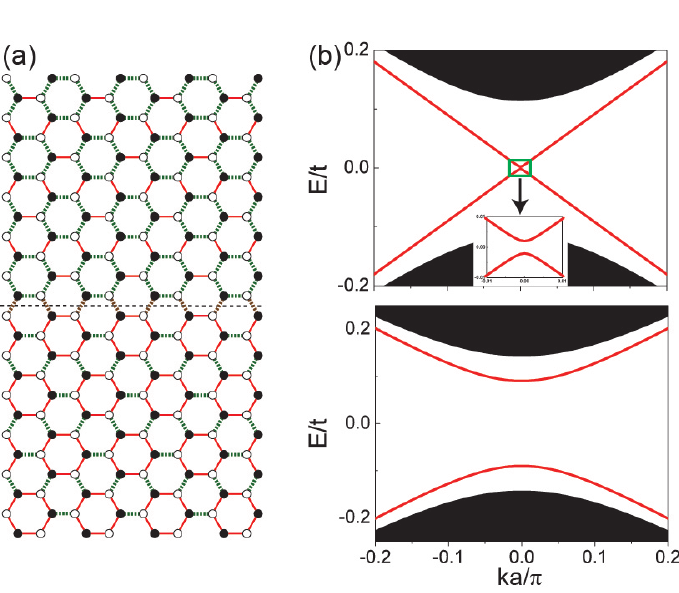}
\caption{(Color online) (a) Schematic plot of a domain wall with $\sqrt{3} \times \sqrt{3}$ lattice deformation. Compared to the Fig. 2 in the main text, the interface here is in the armchair-type. (b), (c) One dimensional energy spectrum of the model. The parameters in the two regions are $(\delta t, \Delta)=(0.06t, ~0.0t)\ /(-0.0675t,~0.0t)$ in (b), and $(\delta t, \Delta)=(0.06t, ~0.09t)\ / (-0.0675t,~0.09t)$ in  (c). The inset of (b) is the zoomed view of the green box. A tiny gap due to intervalley scattering can be clearly observed.
}\label{FigR1}
\end{figure}

\section{Bound energies in topological quantum dots}

\begin{figure}[h]
\centering
\includegraphics[width=16cm]{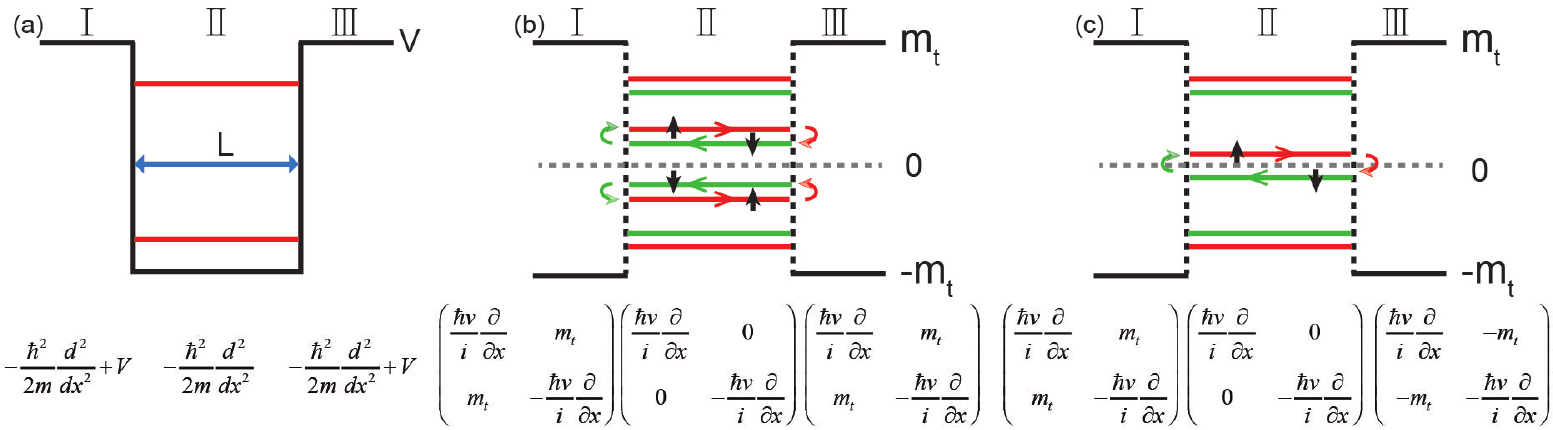}
\caption{(Color online) (a)-(c) Upper panel: schematic plots of three types of 1D quantum potential well with finite-depth. (a) Schr$\mathrm{\ddot{o}}$dinger type. (b), (c) Dirac type. Lower panels: the corresponding effective Hamiltonian in each region. (b), (c) are corresponding to the devices illustrated in Fig. 5(a), (b) in the main text, respectively. In region II of (b), (c), in order to confine a massless Dirac Fermion, its pesudospin direction should rotate by $2\pi$. In (c), the mass term changes its sign between region I and region III, leading to the special Jackiw-Rebbi boundary condition. For simplicity, we assume that $m_t$ is positive in the left side.}\label{FigR1}
\end{figure}

In this section, we derive the bound energies $\varepsilon$ for the two types of quantum dot devices stated in the main text. $\varepsilon$ can be obtained by solving the Dirac-type 1D finite-depth potential well with different boundary conditions [see Fig. S4(b), (c)]. We also derive the bound energies for normal-type 1D finite-depth potential well as a comparison [see Fig. S4(a)].

\noindent In Fig. S4(a), the eigenfunctions in the three regions read:

\begin{eqnarray}
\begin{cases}
\psi_{\rm I}(x) \ \ \ = A e^{\kappa x}, \ \ \ \ \ \  \ \ \ \ \ \ \  \ \ \ x<0  \\
\psi_{\rm II}(x) \ \ = B e^{i k x} +C e^{-i kx}, \ \   0 < x < L  \\
\psi_{\rm III}(x) = D e^{- \kappa x}, \ \  \ \ \ \ \ \ \ \ \ \ \ \ \  x > L
\end{cases}\label{EQ1}
\end{eqnarray}

\noindent where $\kappa =\frac{\sqrt{2m(V-\varepsilon)}}{\hbar}$, $k=\frac{\sqrt{2m\varepsilon}}{\hbar}$. Considering the continuity conditions in the interfaces $\psi_{\rm I}(0)=\psi_{\rm II}(0)$, $\psi_{\rm II}(L)=\psi_{\rm III}(L)$, $\psi_{\rm I}^{'}(0)=\psi_{\rm II}^{'}(0)$ and $\psi_{\rm II}^{'}(L)=\psi_{\rm III}^{'}(L)$, one obtain the equation of the bound energy as:

\begin{equation} \label{EQ2}
e^{2ikL}=\frac{(k+i \kappa)^2}{(k-i \kappa)^2},
\end{equation}

In the condition of $\varepsilon \ll V$, e.g. one dimensional potential well with infinite-depth, Eq. (\ref{EQ2}) reduces:

\begin{equation} \label{EQ3}
e^{2ikL}=1,
\end{equation}

\noindent In Fig. S5(b), the eigenfunctions in the three regions read:

\begin{eqnarray}
\begin{cases}
\psi_{\rm I}(x) \ \ \ = A \left(
                      \begin{array}{c}
                        m_t \\
                        \varepsilon+ i\sqrt{m_t^2-\varepsilon^2} \\
                      \end{array}
                    \right)  e^{\kappa x}
, \ \ \ \ \ \  \ \ \ \ \ \ \  \ \  x<0  \\
\psi_{\rm II}(x) \ \ = B \left(
                       \begin{array}{c}
                         1 \\
                         0 \\
                       \end{array}
                     \right)
 e^{i k x} +C \left(
                \begin{array}{c}
                  0 \\
                  1 \\
                \end{array}
              \right)
  e^{-i kx}, \ \ \ \ \ \ \ \ \ \  0 < x < L  \\
\psi_{\rm III}(x) = D \left(
                      \begin{array}{c}
                        m_t \\
                        \varepsilon - i\sqrt{m_t^2-\varepsilon^2} \\
                      \end{array}
                    \right)  e^{- \kappa x}, \ \  \ \ \ \ \ \ \ \ \ \ \ \ \  x > L
\end{cases}\label{EQ4}
\end{eqnarray}

\noindent where $\kappa =\frac{\sqrt{m_t^2-\varepsilon^2}}{\hbar \upsilon}$, $k=\frac{\varepsilon}{\hbar \upsilon}$. Considering the continuity conditions in the interfaces $\psi_{\rm I}(0)=\psi_{\rm II}(0)$, $\psi_{\rm II}(L)=\psi_{\rm III}(L)$, the bound energy equation reads:

\begin{equation} \label{EQ5}
e^{2ikL}=\frac{\varepsilon + i \sqrt{m_t^2-\varepsilon^2}}{\varepsilon - i \sqrt{m_t^2-\varepsilon^2}},
\end{equation}

We only focus on the low-energy case, in another word $\varepsilon \ll m_t$. Hence the above equation reduces:

\begin{equation} \label{EQ6}
e^{2ikL}=-1.
\end{equation}

\noindent Therefore, the boundary energy are equally spaced since $\varepsilon =\frac{2n+1}{2L}\pi \hbar \upsilon $ ($n$ is an integer). Comparing Eq. (\ref{EQ6}) with Eq. (\ref{EQ3}), it can be observed that there is a $\pi$-phase shift from $x=0$ to $x=L$. This is because the system is described by the massless Dirac equation. The  spinor's two components can be regarded as a pseudospin. To confine a massless Dirac fermion, the pseudospin should change its spin direction twice, as illustrated in Fig. S4(b). The $2\pi$-rotation of the pseudospin leads to the $\pi$ Berry phase.

Next, we derive the bound energy equation for Fig. S4(c). Comparing with Fig. S4(b), the mass term changes its sign in region ${\rm III}$, which introduces the Jackiw-Rebbi boundary condition. The eigenfunctions in these three regions read:

\begin{eqnarray}
\begin{cases}
\psi_{\rm I}(x) \ \ \ = A \left(
                      \begin{array}{c}
                        m_t \\
                        \varepsilon+ i\sqrt{m_t^2-\varepsilon^2} \\
                      \end{array}
                    \right)  e^{\kappa x}
, \ \ \ \ \ \  \ \ \ \ \ \ \  \ \  x<0  \\
\psi_{\rm II}(x) \ \ = B \left(
                       \begin{array}{c}
                         1 \\
                         0 \\
                       \end{array}
                     \right)
 e^{i k x} +C \left(
                \begin{array}{c}
                  0 \\
                  1 \\
                \end{array}
              \right)
  e^{-i kx}, \ \ \ \ \ \ \ \ \ \  0 < x < L  \\
\psi_{\rm III}(x) = D \left(
                      \begin{array}{c}
                        - m_t \\
                        \varepsilon - i\sqrt{m_t^2-\varepsilon^2} \\
                      \end{array}
                    \right)  e^{- \kappa x}, \ \  \ \ \ \ \ \ \ \ \ \  \ \ \ x > L
\end{cases}\label{EQ7}
\end{eqnarray}

\noindent where $\kappa$ and $k$ have the same definition as in Eq. (\ref{EQ4}). By applying the continuity boundary condition at $x=0$ and $x=L$, one obtain

\begin{equation} \label{EQ8}
e^{2ikL + i \pi}=\frac{\varepsilon + i \sqrt{m_t^2-\varepsilon^2}}{\varepsilon - i \sqrt{m_t^2-\varepsilon^2}},
\end{equation}

\noindent with $k=\frac{\varepsilon}{ \hbar \upsilon}$. Comparing Eq. (\ref{EQ8}) with Eq. (\ref{EQ5}), the phase shift of $\pi$ leads to a half-period shift of bound energy $\varepsilon$. Moreover, $\varepsilon =0$ is always a solution of Eq. (\ref{EQ8}) for arbitrary $L$. It verifies that the topological corner state in our case is indeed the celebrated Jackiw-Rebbi zero-mode \cite{Jackiw}.

Moreover, if the mass term of region ${\rm III}$ is changed from $m_t$ to $m_{t'}$, Eq. (\ref{EQ8}) is modified to:

\begin{equation} \label{EQ9}
e^{2ikL + i \pi}=\frac{m_{t'}}{m_t} \frac{\varepsilon + i \sqrt{m_t^2-\varepsilon^2}}{\varepsilon - i \sqrt{m_{t'}^2-\varepsilon^2}}.
\end{equation}

\noindent $\varepsilon =0$ is still the solution to the bound energy equation, which demonstrates the topological nature of the topological corner state.

\section{Electromagnetic Numerical Simulations}
All numerical results described in this section are carried out by the finite element numerical solver (COMSOL Multiphysics). We first obtain the photonic band diagrams of the photonic graphene structure. 2D photonic crystal (PC) is considered by assembling inifinitely long alumina cylinders (with relative permittivity $\varepsilon =7.5$) in air. We consider transverse magnetic polarization with the electric field along the alumina cylinders. 1D band inversion (as shown in Fig. 2 in the main text) occurs either by tuning the intra (inter)-unit-cell separation $h_1$ ($h_2$) between neighboring cylinders in PCs (i.e. lattice deformation) or by manipulating sublattice dielectric cylinder diameters (i.e. sublattice symmetry breaking), which can be characterized by the properties of the calculated photonic band diagrams. In Fig. S5(a), we show the calculated bulk photonic band diagrams of two setups used in the experiments, corresponding to $(-\delta t, \Delta)$ and $(\delta t, -\Delta)$, respectively (the hexagonal unit cell is shown in the inset and the yellow circles represent alumina cylinders). The lattice constant $a=17 \sqrt{3}$ mm and the diameters of cylinders at A and B sublattice sites are $d_0+ \Delta d=6$ mm and $d_0- \Delta d =5$ mm, respectively. The A and B sublattice sites are swapped between the left and the right panel, indicating the change of the sign of $\Delta$. The intra-unit-cell separation is set as $h_1=19/ \sqrt{3}$ mm and $13.5/ \sqrt{3}$ mm in the left and right panel, respectively. Please be reminded of the relationship $2h_1+h_2=a$. The sign of $\delta t$ is reversed by relocating these six cylinders. In both two photonic band diagrams in Fig. S5(a), the 2nd/3rd and the 4th/5th photonic bands are double degenerate at the Brillouin zone center ($\Gamma$ point). The eigenmodes of the PC in left panel can be found in the inset of Fig. 2(g) in the main text as indicated by the grey dashed line. The eigenmodes of the 4th and 5th photonic bands possess a dipole-like symmetry while the lower-frequency doubly-degenerate modes possess a quadrupole-like symmetry, which are the photonic analogues of the $p$ and $d$ electronic orbits, respectively. The frequencies of the dipole ($p$-orbital) modes are lower than the quadrupole one ($d$-orbital) for the PC in the right panel, indicating a band inversion induced by lattice deformation. Even in the condition of $\Delta d=0$ (all cylinders of the same diameter), such a 1D phase transition still occurs as shown in Fig. S5(b), which is identical to the scenario as discussed in Ref. \cite{ZHHang}. When $h_1=h_2$, the structure is identical to graphene.

\begin{figure}[h]
\centering
\includegraphics[width=14cm]{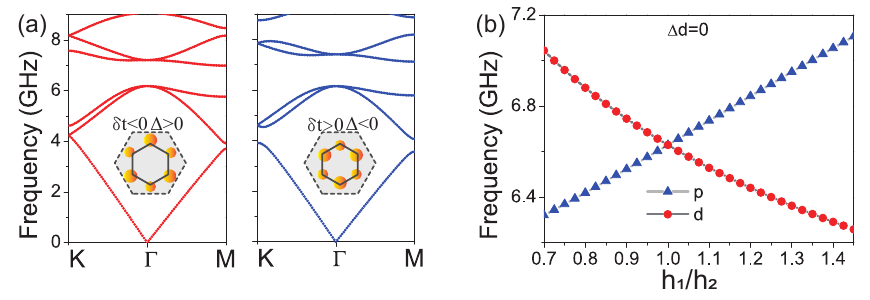}
\caption{(Color online) (a) Left and right panel: the calculated photonic band diagrams of 2D PCs with $(-\delta t, \Delta )$ and $(\delta t, -\Delta)$ as adopted in the experiments, respectively. (b) 1D phase transition occurs with $\Delta d=0 $.}\label{FigS5}
\end{figure}

The photonic dispersion of the topological kink states in Region III [Fig. S6(a)] can be obtained by calculating a 2D supercell [Fig. S6(b)], where five layers of PC with $h_1=19/\sqrt{3}$ mm ($13.5/ \sqrt{3}$ mm) are assembled on the left (right). Periodic boundary conditions are applied in both directions of the supercell. The cylinder diameter at site A (B) is 5 mm (6 mm) in both PC structures, which  corresponds to a $\sqrt{3} \times \sqrt{3}$ domain wall. Two dispersions (in black lines) emerge in the gap of the bulk photonic bands (denoted by grey shadowed regions) and a eigenfield distribution confining on the interface (domain wall) as shown in Fig. S6(b) clearly indicates its kink state properties. In principle, the photonic graphene system we study is a 2D system. However, in order to facilitate the automatic electric field mapping procedure to obtain the field distributions as shown in the main text, a thin layer (0.5 mm) of air gap is allowed between the alumina cylinders and the top metallic plate of the parallel waveguide, which makes it a 3D system. We calculated a corresponding 3D supercell by considering the air gap. The results on three domain walls I, II and III as discussed in the main text are shown in the left, middle and right panels of Fig. S6(c). The numerical analysis manifests that such a thin air gap only increases the frequencies of the topological states while their physical properties remain unchanged. The band diagram in Fig. S6(d) is obtained by combining three panels of Fig. S6(c), which is similar to Fig. 3(a) in the main text. We also obtain the electric field distributions of the topological kink states and topological corner states by numerically simulating the experimental setup shown in Fig. 1(a). Corresponding simulation results are shown in Fig. S6(e)-(h). Good consistence between numerical simulations and experiments (Fig. 3 in the main text) are found.

\begin{figure}[h]
\centering
\includegraphics[width=14cm]{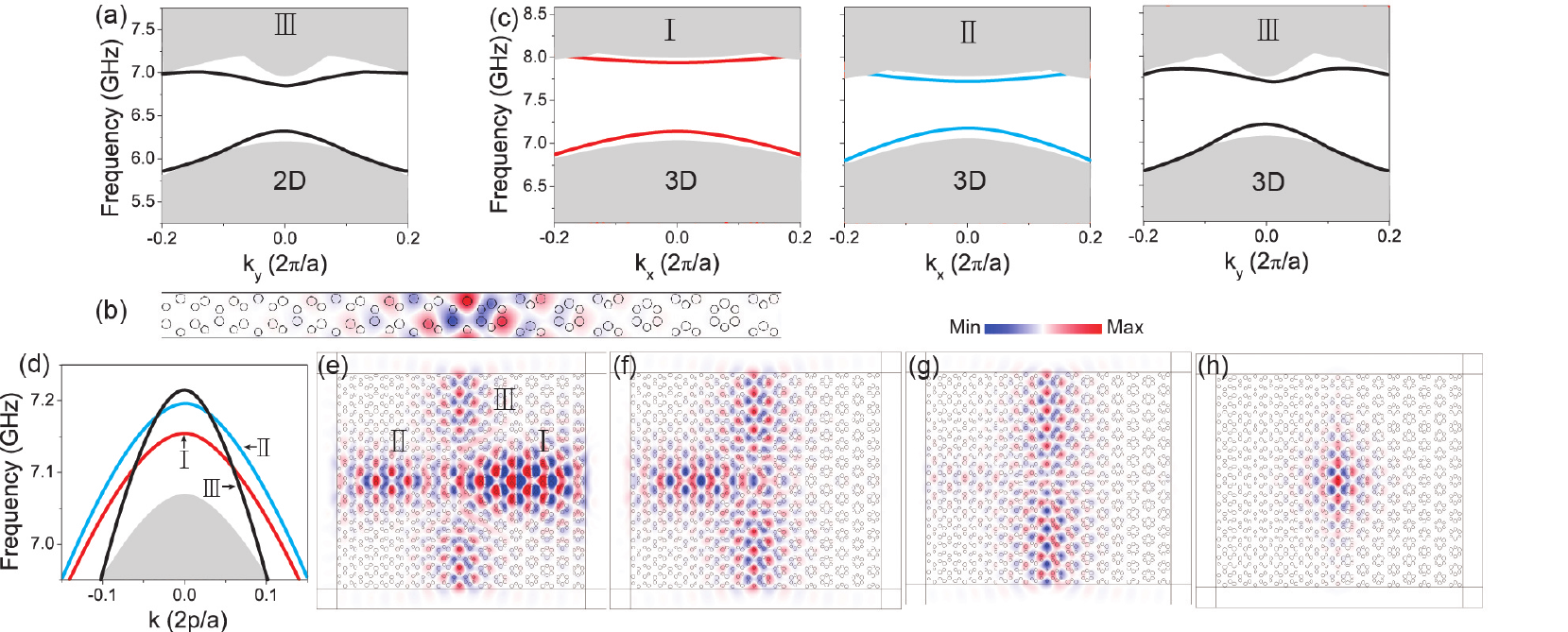}
\caption{(Color online) (a) The calculated band diagram of 2D PCs. (b) The eigenmode of the topological kink state. (c) The 3D projected bands in region I, II and III, which are shown in the left, middle and right panels, respectively. (d) The combination of the band diagrams in the three panels of (c). (e)-(h) 2D simulations of electric field distribution of the experimental setup at frequencies 6.22, 6.27,  6.29 and 6.523~GHz, respectively.}\label{FigS6}
\end{figure}

\section{Microwave Experimental Setup}
In this section, we describe the microwave experiments in details. A home-made parallel-plate waveguide microwave field mapping system is used to characterize the properties of the topological states in our microwave PC system. The positions of alumina cylinders assembled on the bottom metallic plate can be precisely controlled because of the centimeter length scale. Absorbing materials [blue regions in Fig. 1(a) in the main text] is adopted to avoid unexpected scattering. An emitting antenna at the star position in Fig. 1(a) in the main text is used to excite the topological states while a probe antenna through the upper metallic plate of the waveguide is used to measure the local electric field. Both the emitting and probe antennas are connected to a vector network analyzer (VNA, Agilent E5071C) for data acquisitions. The bottom metallic plate, along with the samples and emitting antenna is mounted on a computer-controlled translational stage and thus the time-harmonic $E_z$ field distribution ($z$ is the direction parallel to the cylinder axis) at different frequencies can be obtained by point-by-point recording. The alumina cylinders are made by a commercial vendor and their height is 8 mm. In order to facilitate the movement of the translational stage, the separation between the two parallel metallic plates is 8.5 mm, which is the origin of the thin air gap (0.5 mm) discussed in the previous section. Besides the field distributions at different frequencies, we also integrate the electric field magnitude of the whole electric field distributions and the corresponding result is shown in Fig. S7(a), where the topological corner state (black star) can be clearly observed. The 2D counterpart is computed as shown in Fig. S7(b) and a symmetry Lorentzian spectrum line shape is observed. Please be noted that the corner mode is at a lower frequency in the 2D case. The quality factor of the topological corner state, as defined as the center frequency dividing the full width at half maximum is also obtained as $Q=3.4 \times 10^4$. Furthermore, we also analyze the confinement of the topological corner states. From the local field distributions and the line plots along the dashed lines as shown in Fig. S7(c), (d), the decay lengths are 27.63 mm (15.73 mm)  and 26.34 mm (27.02 mm)  towards $-x$ ($x$) and $-y$ ($y$) direction, respectively. It indicates a good confinement of the topological corner mode near $x=y=0$.

\begin{figure}[h]
\centering
\includegraphics[width=14cm]{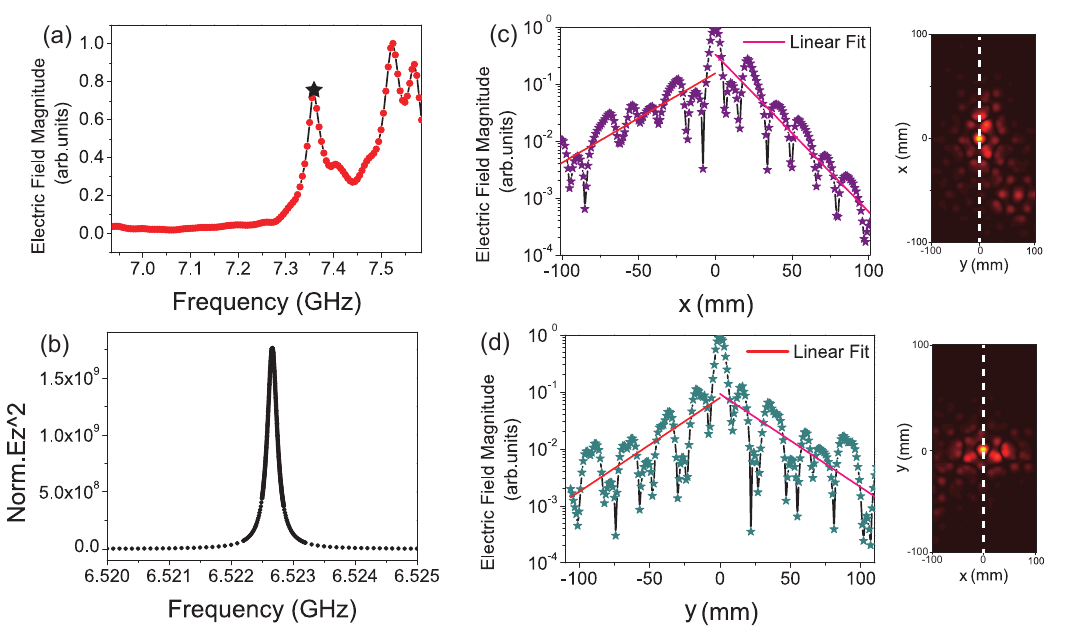}
\caption{(Color online) (a) The measured electric field magnitude integrated at different frequencies. The black star indicates the frequency of the topological corner state. (b) the calculated spectrum line shape and $Q$ factor of the corner state. (c), (d) The local confinement of the corner state along the $x$ and $y$ directions, respectively.}\label{FigS7}
\end{figure}

\begin{figure}[h]
\centering
\includegraphics[width=14cm]{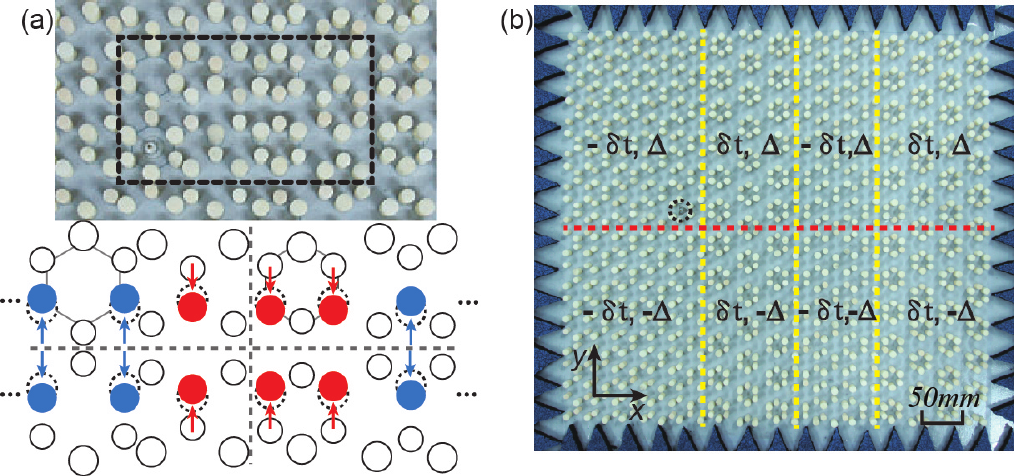}
\caption{(Color online) (a) The photograph and the schematic plot of the random dislocations of the alumina cylinders introduced near the intersection region of the domain walls. (b) The photograph of the experimental setup with coupling between topological corner states. The AB-BA ($\sqrt{3} \times \sqrt{3}$) domain wall is denoted by red (yellow) dashed line.}\label{FigS7}
\end{figure}

Even in the condition that the alumina cylinders are randomly positioned, the topological corner state still survive. The measured electric field distribution of the robust corner state is shown in Fig. 3(h) in the main text. Figure S8(a) displays the positions of alumina cylinders close to the intersection region of the domain walls. The dotted circles represent the original positions  without random dislocation. PC alumina cylinders move towards (away from) the AB-BA domain (shown in dashed line) are shown in red circles (blue circles). The photograph of the experimental setup with coupling between topological corner states (Fig. 4 in the main text) is shown in Fig. S8(b).

\end{widetext}

\end{document}